\documentstyle[12pt]{article}
\title{\bf On  gauge transformations of B\"acklund type and higher
order nonlinear Schr\"odinger equations}
\author{ 
\bf     Gerald A. Goldin\footnote{e-mail: gagoldin@dimacs.rutgers.edu}
\\
\it   Departments of Mathematics and Physics, Rutgers University, \\
\it  Busch Campus, Piscataway, New Jersey 08854\\
\bf  Vladimir M. Shtelen\footnote{e-mail: shtelen@math.rutgers.edu}\\
\it   Department of Mathematics, Rutgers University, Busch Campus,\\
\it   Piscataway, New Jersey 08854
}

\date{\today}
\begin{document}
\setcounter{section}{0}
\renewcommand{\theequation}{\arabic{section}.\arabic{equation}}
\renewcommand{\vec}{\bf}

\maketitle

\begin{abstract}
We introduce a new, more general type of nonlinear gauge transformation in 
nonrelativistic quantum mechanics that 
involves derivatives of the wave function and  belongs
to the  class of B\"acklund transformations. These transformations satisfy
certain reasonable, previously proposed requirements for  gauge transformations.
Their application to the  Schr\"odinger equation 
results in higher order partial differential equations.
As an example,
we derive a  general family  of 6th-order
 nonlinear  Schr\"odinger equations, closed under our nonlinear gauge
group. We also introduce a new gauge invariant current 
${\bf \sigma}=\rho {\bf \nabla}\triangle \ln \rho $, where $\rho=\bar\psi \psi$. We 
 derive gauge invariant quantities,  and characterize the subclass of the 6th-order equations
that is gauge equivalent to the free
Schr\"odinger equation. We relate our development to nonlinear equations 
studied by Doebner and Goldin, and by Puszkarz.
\end{abstract} 

\newpage
\section{Introduction}
 The notion of nonlinear gauge transformation, introduced in quantum mechanics by 
Doebner and Goldin,
extends the usual group of unitary gauge transformations.$^{1-3}$ The resulting
nonlinear transformations act on a parameterized family of nonlinear Schr\"odinger equations 
(NLSEs) that includes the linear Schr\"odinger equation as a special case. They are called gauge 
transformations because they leave invariant the outcomes of all physical measurements.
In this paper we extend the notion of   gauge transformation further to include
transformations that depend explicitly on derivatives of the wave function.
The result is a group of transformations of   B\"acklund type.$^4$

As described in earlier work,$^3$ a (nonlinear) gauge transformation is 
implemented by a transformation $\psi' = {\mathcal{N}}[\psi]$, assumed to satisfy the following conditions:
\vspace{\smallskipamount}
\begin{itemize}
\item 1. The  {\it  principle of gauge-independence of positional
measurements:} Invariance is required of all quantities describing  outcomes of positional measurements, 
including {\it sequences} of measurements performed successively at different times. In particular,
$\rho({\bf x},t)=|\psi({\bf x},t)|^2$ should be invariant under $\mathcal{N}$  for the single-particle 
wave function $\psi$.
\item 2. {\it Strict locality:} If $\psi$ is  a single-particle function, the value 
of $\psi'$ at $({\bf x},t)$ is assumed to depend 
 only on the value of ${\bf x}$, the value of $t$, and the value of $\psi$ at $({\bf x},t)$.
\item 3. A {\it separation condition:} If $\psi^{(N)}$ is a wave function describing a set of $N$ 
noninteracting particles (i.e., a product state), then $\psi^{(N)'}$ is well defined as 
the product of gauge transformed single particle wave functions. This condition ensures that 
gauge transformations extend to the whole $N$-particle hierarchy of wave functions in a way that
subsystems that are uncorrelated  remain so in the gauge-transformed theory.
\end{itemize}
Here we modify the condition of strict locality, allowing $\psi'({\bf x},t)$ to depend not 
only on the values of $\psi({\bf x},t)$, ${\bf x}$, and  $t$, 
but also on finitely many spatial derivatives 
of $\psi$ evaluated at $({\bf x}, t)$. Thus our transformations are local, in that $\psi'({\bf x},t)$ 
does not depend on space-time points any distance from $({\bf x}, t)$, but they are no longer  ``strictly"
local, since derivative terms are allowed. We shall call this property {\it weak locality}. 
One motivation for introducing this generalization  
is to explore the relation between the resulting nonlinear gauge
generalization of 
the Schr\"odinger equation  
and the equations proposed by Puszkarz.$^5$

The condition that our set of transformations forms a group (i.e., that it is closed under composition
and includes all inverse transformations) while the number of derivatives of $\psi$ remains bounded, 
imposes an additional restriction. This {\it group property} is automatically satisfied in the 
strictly local theory, but
here it requires explicit discussion. Thus, we shall add it to the   conditions already mentioned. 
We then call the transformations that obey the 
following four conditions  {\it weakly local  gauge transformations}:
1. the principle of gauge-independence of positional measurements;
$2'$. weak locality; 
3. the separation condition; and
4. the group property.

In Sec. 2 of this paper, we first consider a general class of nonlinear, single particle 
Schr\"odinger equations 
that are equivalent to the  free  Schr\"odinger equation  
under the assumption that condition 1 is satisfied. Using  the other three conditions, 
we obtain a particularly simple form for weakly local gauge transformations.
Following the method of ``gauge generalization,"$^3$ we  then derive a  general family  of 6th-order
nonlinear  Schr\"odinger equations, closed under our nonlinear gauge
group, which are  not all equivalent to the free  2nd-order Schr\"odinger  
equation. In Sec. 3 we construct a complete set of 
gauge invariant quantities. As  particular cases, we  use these to characterize the  subclass 
of the 6th-order equations that are  gauge equivalent to the  
Schr\"odinger equation, and those equivalent to the  wider class of nonlinear equations 
studied by Doebner and Goldin.
We further relate our development to the nonlinear equations proposed by Puszkarz
based on additional quantum currents that involve higher derivatives of $\psi$.

\section{Gauge Transformations and NLSEs}
\setcounter{equation}{0}
Consider the transformation 
\begin{equation}
\psi'({\bf x},t)\,=\,e^{i\varphi}\psi({\bf x},t),
\end{equation}
where $\varphi$ is a real-valued functional that depends on $\psi, {\bf x}$, and $t$.
By this we mean that $\varphi$ can depend explicitly on $\psi$, $\bar \psi$, derivatives
of $\psi$ and $\bar \psi$ of arbitrary order, integrals or integral transforms
of $\psi$ and $\bar \psi$, etc., as well as directly on ${\bf x}$
and $t$. Eq. (2.1) 
preserves the probability density  
$\rho({\bf x},t)=\bar\psi({\bf x},t)\psi({\bf x},t)$, as required by the first condition in Sec. 1,
but if nonlocal it does not generally respect sequences of positional measurements.
The following then describes the general class of NLSEs that are
equivalent via (2.1) to the free Schr\"odinger equation:
if $\psi'$  
satisfies
\begin{equation}
   i\frac{\partial\psi'}{\partial t} + \frac{\hbar}{2m} \triangle \psi'\,=\,
i\frac{\partial\psi'}{\partial t} - \nu_1' \triangle \psi'=0, 
\end{equation}
then $\psi$ satisfies the NLSE
\begin{equation}
 i\frac{\partial\psi}{\partial t}    -\nu_1' \triangle \psi + iI[\psi, {\bf x}, t] \psi + 
R[\psi, {\bf x}, t] \psi=0,                                                           
\end{equation}
where
\begin{equation}
R[\psi, {\bf x}, t]\,=\,\frac{\partial\varphi}{\partial
t}-2\nu_1'(\frac{{\bf \nabla} \varphi\cdot\hat{ \bf j}}
{\rho}+\frac{1}{2}({\bf \nabla}\varphi)^2)
\end{equation}
and
\begin{equation}
I[\psi, {\bf x}, t] \,=\, \nu_1' (\triangle \varphi+\frac{{\bf \nabla}
\varphi\cdot{\bf \nabla}\rho}{\rho})\,=\,
\nu_1'[\frac{1}{\rho}({\bf \nabla}\cdot(\rho{\bf \nabla}\varphi)],
\end{equation}
with
\begin{equation}
\hat{ {\bf j}}=\frac{m}{\hbar}{\bf j}=\frac{1}{2i}[\bar\psi{\bf\nabla}\psi-({\bf \nabla}\bar\psi)\psi].
\end{equation}
The verification is by direct substitution of (2.1) into (2.2).

As was shown by Doebner and Goldin$^1$, a general form for strictly local
 gauge transformations (that satisfy all  the initial
requirements discussed in Sec. 1) corresponds to the choice
\begin{equation}
\varphi=\frac{1}{2}\gamma(t)\ln\rho + [\Lambda(t) -1]S
+ \theta({\bf x},t), \quad  \Lambda \neq 0 ,
\end{equation} 
where $\psi\,=\, \sqrt{\rho}\,e^{iS}$. For simplicity, we
consider $\theta({\bf x},t) \equiv 0$.
The family of NLSEs with arbitrary coefficients that directly generalizes (2.3) 
and is invariant (as a family) under gauge transformations (2.1)
with $\varphi$ as in (2.7), 
then has the form$^1$
\begin{equation}
i\frac{\partial\psi}{\partial t}\,=\,\{i\sum_{j=1}^{2}\nu_j(t)R_j+\sum_{j=1}^{5}\mu_j(t)R_j\}\psi, 
\end{equation}
where 
\begin{equation}
R_1=\frac{{\bf \nabla}\cdot \hat {\bf j}}{\rho}, \quad
R_2=\frac{\triangle \rho}{\rho}, \quad R_3=\frac{{\hat {\bf
j}}^2}{\rho^2}, \quad
R_4=\frac{{\hat {\bf j}} \cdot {\bf \nabla}\rho}{\rho^2}, \quad R_5=\frac{{({\bf \nabla} \rho)}^2}{\rho^2}.
\end{equation}
In obtaining (2.8), one uses the identity $\triangle\psi/\psi\,=\,iR_1 + \frac{1}{2}R_2
- R_3 - \frac{1}{4}R_5$.
Invariance of the family (2.8) under (2.1) and (2.7) means that if $\psi$
satisfies an equation in this family with coefficients
$\nu_j$ and $\mu_j$, then $\psi'$ satisfies another equation in the family
with coefficients $\nu_j'$ and $\mu_j'$; thus our choice of the primed
coefficient $\nu_1'$ in
writing Eq. (2.2).

Now the class of nonlinear gauge transformations in quantum mechanics can be essentially extended if
we replace strict locality by weak locality, thus allowing the gauge functional 
$\varphi$ to depend on derivatives of $\psi$.
Under this assumption the gauge transformation is no longer simply a point
transformation; it is a {\it  B\"acklund transformation}.$^4$ 
Here we consider gauge transformations of B\"acklund type that form a
group, satisfying the physically motivated
requirements discussed in Sec. 1, with strict locality  replaced by weak locality.

We observe that if $\varphi$ is permitted to depend on derivatives of $S$ as well as
derivatives of $\rho$, then the set of gauge transformations in general does not respect the 
group property. However, if the derivatives of $S$ are excluded from  $\varphi$, then
the transformations do respect this property.
One way to see this
is to write nonlinear gauge transformations as they act on logarithmic
coordinates $T$ and $S$, with $\ln \psi = T + iS$ (so that $T = \frac{1}{2} \ln \rho)$,
omitting for simplicity the explicit ${\bf x}$ and $t$ dependence:
\begin{equation}
\mathbf{}\left(\begin{array}{c}S'\\T'\end{array}\right)=
\mathbf{}\left(\begin{array}{cc}L&G\\0&1\end{array}\right)
\mathbf{}\left(\begin{array}{c}S\\T\end{array}\right),
\end{equation}
where $L$ is a linear or nonlinear functional of $S$ and its derivatives,
and $G$ is a linear or nonlinear functional of $T$ and its derivatives.
In the strictly local case, we have $L[S] = \Lambda S$ and $F[T] = \gamma \,T$.
If we perform two transformations (2.10) successively, $T'' = T' = T$
and $S'' = L_2[L_1[S] + G_1[T]\,] + G_2[T]$. Then derivatives present
in the form of $G$ never act successively, so that their order
does not increase; but derivatives in the form of $L$ do act successively.
Thus the group property, with the condition that the number of derivatives
of $\psi$ remains bounded, rules out derivative terms in $L$---but not in $G$.

Now a simple gauge transformation that is no longer strictly local,
but satisfies the four requirements discussed in Sec. 1, has the form (2.1) with 
\begin{equation}
\varphi\,=\,\frac{1}{2}\gamma\ln\rho + (\Lambda -1) S+\eta\triangle\ln\rho\,=\,
\frac{1}{2}\gamma\ln\rho+(\Lambda -1) S +\eta(R_2-R_5),
\end{equation}
where $\eta$ is a real parameter that, like $\gamma$ and $\Lambda$,
can in principle depend on $t$.
This corresponds to the choice $G[T] = \gamma\,T + \eta\triangle T$ in (2.10).
Thus we have a group of nonlinear gauge transformations
modeled on three (in general time-dependent) parameters,
obeying the group law
\begin{equation}
{\mathcal N}_{(\gamma_2,\Lambda_2,\eta_2)} \circ {\mathcal N}_{(\gamma_1,\Lambda_1,\eta_1)}
= {\mathcal N}_{(\gamma_2 + \Lambda_2\gamma_1, \Lambda_2\Lambda_1, \eta_2 + \Lambda_2\eta_1)}.
\end{equation}
But we note further that $G[T]$ need not be linear in $T$. Indeed,
while the linear term $\triangle \ln \rho \,=\, R_2-R_5$ satisfies the
separation condition, its nonlinear parts $R_2$ and $R_5$ do so separately!
Considering a two-particle product wave function $\psi^{(2)}({\bf x_1}, \ {\bf x_2}, \ t)\,=\,
\psi_1({\bf x_1}, t)\psi_2({\bf x_2}, t)$, and defining 
$\rho^{(2)}=\overline{\psi^{(2)}}\psi^{(2)}$,  
$\rho_1=\bar\psi_1\psi_1$, and $\rho_2=\bar\psi_2\psi_2$, we have 
$$
R^{(2)}_2[\psi^{(2)}]= \frac{\triangle^{(2)}\rho^{(2)}}{\rho^{(2)}}=
\frac{\triangle^{(2)} (\rho_1\rho_2)}{\rho_1\rho_2}= \frac{\triangle_1\rho_1}
{\rho_1} \frac{\triangle_2 \rho_2}{\rho_2}= R_2[\psi_1]R_2[\psi_2],
$$ 
where $\triangle^{(2)} =\triangle_1 + \triangle_2$. Similarly for $R_5$:
$$
R^{(2)}_5 [\psi^{(2)}]=\frac{[{\bf
\nabla}^{(2)}\rho^{(2)}]^2}{{(\rho^{(2)})}^2}=
\frac{[({\bf \nabla}_1, 
{\bf \nabla}_2)\rho_1\rho_2]^2}{(\rho_1\rho_2)^2}=R_5[\psi_1]R_5[\psi_2].
$$
Thus a further generalization of (2.11) that gives weakly local
nonlinear gauge transformations is to 
allow the derivative terms to enter with different coefficients:
\begin{equation}
\varphi\,=\,\frac{1}{2}\gamma\ln\rho + (\Lambda -1) S +
\eta_1 R_2 + \eta_2 R_5.
\end{equation}

Let us next write the gauge generalized family of NLSEs
derived from (2.11).
Beginning with the standard, free Schr\"odinger equation in the form
\begin{equation}
i\frac{\partial \psi'}{\partial t}\,=\,-\frac{\hbar}{2m}[iR_1' +(\frac{1}{2}R_2' -R_3' -\frac{1}{4}R_5')] \psi',
\end{equation}
where $R'_j$ means $R_j[\psi']$,
we transform by (2.1) with $\varphi$ as in (2.11), and from (2.3)-(2.5)
we find the form of the resulting NLSEs for $\psi$. We generalize, following Ref. 3, by
allowing arbitrary coefficients for the nonlinear functionals,
maintaining the invariance of the family of NLSEs
under the nonlinear gauge group.
In this fashion, we obtain the following equations:
\begin{equation}
i\frac{\partial\psi}{\partial t}\,=\,\{i\sum_{j=1,2,6}\nu_jR_j+\sum_{j=1}^{12}\mu_jR_j\}\psi
\,=\, \{i\hat I + \hat R\}\psi, 
\end{equation}
where $R_1,...,R_5$ are as in (2.9), and where the
new functionals $R_6,...,R_{12}$ are given by:
 
\begin{equation}
R_6=\frac{{\bf \nabla}\cdot {\bf \sigma}}{\rho},\qquad
R_7=\frac{{\hat {\bf j}} \cdot {\bf \sigma}}{\rho^2},\qquad 
R_8=\frac{{\bf \sigma} \cdot {\bf \nabla}\rho}{\rho^2},
\end{equation}

$$
R_9=\frac{{ {\bf \sigma}}^2}{\rho^2}, \quad
R_{10}=\triangle R_1,\quad
R_{11}=\triangle R_2,\quad
R_{12}=\triangle R_6,
$$
with 
\begin{equation}
{\bf \sigma}=\rho {\bf \nabla}\triangle \ln \rho =
\rho {\bf \nabla} (R_2-R_5).
\end{equation}
Note that the functionals $R_6,...,R_{11}$ involve no higher than fourth derivatives
of $\psi$, but the presence of the term $R_{12}$ in (2.15) makes it in general
of 6th order. If we use (2.13) in place of (2.11), we shall need
separately the new currents $\rho{\bf \nabla}R_2$ and $\rho{\bf
\nabla}R_5$. These give rise to additional nonlinear functionals in $\psi$.

Equation (2.15) still conserves the quantum probability $\bar\psi\psi$
It gives rise to the gauge invariant current
\begin{equation}
{\bf J}^{gi}=-2(\nu_1\hat{{\bf j}}+\nu_2{\bf \nabla}\rho +\nu_6{\bf \sigma})
\end{equation}
that enters the continuity equation
\begin{equation}
\frac{\partial\rho}{\partial t} =-{\bf \nabla}\cdot {\bf J}^{gi}= 2\hat I\rho.
\end{equation}


\section{Gauge transformations and invariants for the family of
6th-order NLSEs}
 \setcounter{equation}{0}

Under the gauge transformations (2.1), with $\varphi$ given by (2.11)
the coefficients $\nu_j ,\  \mu_j$ of (2.15) transform as follows:
\begin{equation}
\nu_1'=\frac{\nu_1}{\Lambda},\qquad 
\nu_2'=\nu_2-\frac{1}{2}\gamma\frac{\nu_1}{\Lambda},\qquad
\nu_6'=\nu_6-\frac{\eta}{\nu_1\Lambda}\qquad (\Lambda=\lambda+1);
\end{equation}
\begin{equation}
\mu_1'=\mu_1 - \frac{\gamma\nu_1}{\Lambda}, \qquad
\mu_2'=\Lambda\mu_2 -\frac{1}{2}\gamma\mu_1
+\frac{\gamma^2}{2\Lambda}\nu_1 -\gamma\nu_2, \qquad
\mu_3'=\frac{\mu_3}{\Lambda}
\end{equation}
$$
\mu_4'=\mu_4 - \frac{\gamma \mu_3}{\Lambda}, \qquad
\mu_5'=\Lambda\mu_5 -\frac{1}{2}\gamma\mu_4
+\frac{\gamma^2}{4\Lambda}\mu_3 , 
$$
$$
\mu_6'=\Lambda\mu_6 -\gamma\nu_6
-\eta\mu_1 + \frac{\eta \gamma}{\Lambda}\nu_1, \qquad
\mu_7'=\mu_7 - \frac{2\eta \mu_3}{\Lambda} 
$$
$$
\mu_8'=\Lambda\mu_8 -\eta \mu_4 - \frac{1}{2}\gamma\mu_7 + \frac{\gamma\eta \mu_3}{\Lambda},\qquad
\mu_9'=\Lambda\mu_9 -\eta \mu_7 + \frac{\eta^2 \mu_3}{\Lambda},\qquad 
\mu_{10}'=\mu_{10} - \frac{2\eta \nu_1}{\Lambda},
$$
$$
\mu_{11}'=\Lambda\mu_{11}- 2\eta \nu_2 -\frac{1}{2}\gamma\mu_{10} + \frac{\gamma\eta \nu_1}{\Lambda},\qquad
\mu_{12}'=\Lambda\mu_{12}- 2\eta \nu_6 -\eta \mu_{10}+ \frac{2\eta^2 \nu_1}{\Lambda}.
$$
Note that as expected, $\eta$ does not enter
the transformation laws for $\nu_1$, $\nu_2$, or $\mu_1, ..., \mu_5$, which are the same as in Refs. 1-3.
Note also that if we begin with $\mu_{12} = 0$, then $\eta \not= 0$
leads to $\mu_{12}' \not= 0$; thus we cannot have an invariant
family of 4th-order partial differential equations for these
transformations.

We now write functionally independent gauge invariants $\tau_j \
(j=1,2,...,12)$ as follows:
\begin{equation}
\tau_1= \nu_2-\frac{\mu_1}{2},\,\,\, 
\tau_2=\nu_1\mu_2-\mu_1\nu_2,\,\,\,
\tau_3=\frac{\mu_3}{\nu_1},\,\,\,
\tau_4=\mu_4-\mu_1\frac{\mu_3}{\nu_1},\,\,\,
\hat\tau_5=\mu_5\mu_3-(1/4)\mu_4^2,
\end{equation}
$$
\tau_6=\mu_6\nu_1-\mu_1\nu_6,\qquad
\tau_7=\mu_7-2\nu_6\frac{\mu_3}{\nu_1}, \qquad
\tau_8=\mu_8\nu_1-\mu_4\nu_6+\mu_6\mu_3-(1/2)\mu_7\mu_1, \qquad
$$
$$
\tau_9=\mu_9\mu_3-(1/4)\mu_7^2, \,\,\,
\tau_{10}=\mu_{10}-2\nu_6, \,\,\,
\tau_{11}=\mu_{11}\nu_1-\mu_{10}\nu_2, \,\,\,
\tau_{12}=\mu_{12}\nu_1-\nu_6^2-(1/4)\mu_{10}^2.
$$
In this list of gauge invariants, we have
included a new quantity $\hat\tau_5$ instead of the original 
$\tau_5=\nu_1\mu_5-\nu_2\mu_4+ \nu_2^2(\mu_3/\nu_1)$ that was used
in Refs. 1-3, 
since the expression
for $\hat\tau_5$ is simpler.
The relation between these two
gauge invariants is, of course,
wholly gauge invariant:
$\hat\tau_5=\tau_3\tau_5 +\tau_1\tau_3(\tau_4- \tau_1\tau_3)-(1/4)\tau_4^2= 
\tau_3\tau_5 - (\tau_1\tau_3-\frac{1}{2}\tau_4)^2$.

It should be noted that (2.15) is invariant under Galilean transformations
\begin{equation}
{\tilde{\bf x}}={\bf x} - {\bf v}t, \ \tilde{t}=t, \
\tilde{\psi}({\tilde{\bf x}}, \tilde{t})\,=\,\psi({\bf x}, t)\,e^{\,\frac{i}{2\nu_1}({\bf x}\cdot{\bf v} +
\frac{1}{2}v^2t)} 
\end{equation}
when 
\begin{equation}
\frac{\mu_3}{\nu_1}=-1, \   \mu_1+\mu_4=0, \    \mu_7+\mu_{10}=0,
\end{equation}
and consequently, the gauge invariants $\tau_1,$... $\tau_{12}$ must satisfy the conditions 
\begin{equation}
\tau_3=-1, \quad \tau_4=0, \quad \tau_7 + \tau_{10}=0.
\end{equation}
Under time reversal, all the coefficients $\nu_j, \mu_j$
change sign. Thus time reversal invariance requires
\begin{equation}
\tau_1 = 0, \quad  \tau_4 = 0,  \quad  \tau_7 = 0,  \quad \tau_{10} = 0.
\end{equation}
In particular, when (2.15) is the 
Schr\"odinger equation, we have
\begin{equation}
\nu_1=-\frac{\hbar}{2m},\quad  \mu_2=-\frac{\hbar}{4m}, \quad  \mu_3=\frac{\hbar}{2m}, \quad   
\mu_5=\frac{\hbar}{8m},
\end{equation}
and all other coefficients are zero. Eqs. (3.7) then give
\begin{equation}
\tau_2=\frac{\hbar^2}{8m^2}, \quad \tau_3=-1, \quad \tau_5=\frac{\hbar^2}{16m^2}, 
\end{equation}
with all other $\tau$'s equal to zero. For the equations
studied by Doebner and Goldin, $\tau_1, ..., \tau_5$ are arbitrary,
but $\tau_6, ..., \tau_{12}$ are zero.

Some of the equations discussed by Puszkarz,$^5$ 
belong to the class (2.15), when $\mu_{12}=0$.
Puszkarz's modification of the Schr\"odinger equation
is the formal extension of the equations
of Doebner and Goldin obtained by modifying  
the current (2.6), adding to it any or all
of the following terms with higher derivatives:
$$\rho \triangle (\frac{{\bf j}}{\rho}),  \quad   
\rho {\bf \nabla} (\frac{{\bf j}\cdot {\bf \nabla}\rho}{\rho^2}), \quad   
\rho {\bf \nabla} (\frac{{\bf j}^2}{\rho^2}), \quad   
\rho {\bf \nabla} R_2, \quad  \rho {\bf \nabla}R_5.$$ 
Since  Puszkarz's modification directly affects only the
imaginary part 
of the nonlinear functional for $i\frac{\partial \psi}{\partial t} / \psi$,
namely $(-1/2\rho){\bf \nabla}\cdot{\bf J}$ where $\bf J$ is the current
that appears in the equation of continuity,
and does not  change the real part,  the 
resulting equation is 4th-order. Our equations are
in general 6th-order because 
of the term with $R_{12}$,
which is needed in order to maintain invariance under
the nonlinear gauge group. The equations of Puszkarz with the first three currents 
do not belong to any family that is closed under
a group of weakly local nonlinear gauge transformations,
since the transformations 
giving rise to those currents involve derivatives of the phase $S$. His equations
with the latter two currents belong to the family obtained from (2.15) through
gauge  generalization.

In short, we have obtained a natural family of 6th-order partial
differential equations invariant (as a family) for nonlinear gauge transformations
of B\"acklund type, that includes a subclass gauge equivalent to the linear
Schr\"odinger equation, a wider subclass gauge equivalent to the
equations that Doebner and Goldin studied, and another subclass that
intersects the family of equations proposed by Puszkarz. Given a particular
equation in our family, we can calculate the $12$ gauge-invariant parameters,
and from these immediately determine whether the equation is physically
equivalent to the free Schr\"odinger equation or an equation of
Doebner-Goldin type, and whether it is Galilean and/or time-reversal
invariant.


\newpage
\begin{thebibliography}{}
%
\bibitem{DG96}H.-D.~Doebner and G.~A.~Goldin,
 {Introducing nonlinear gauge
 transformations in
 a family of nonlinear Schr\"odinger equations}, 
Phys.~Rev.~A~{\bf 54}, 3764 (1996).
%
\bibitem{G97} G.~A.~Goldin,
 {Gauge transformations for a family of
 nonlinear Schr\"odinger  equations}, 
J.~Nonl.~Math.~Phys.~{\bf 4}, 6 (1997).
%
\bibitem{DGN99} H.-D.~Doebner,
G.~A.~Goldin, and P.~Nattermann,
 {Gauge transformations in quantum mechanics
 and the unification of nonlinear Schr\"odinger
 equations},  
J.~Math.~Phys.~{\bf 40}, 49 (1999).
%
\bibitem{AI} R.~L.~Anderson and N.~Kh.~Ibragimov,
{Lie-B\"acklund Transformations in Applications},
SIAM, Philadelphia (1979).
%
\bibitem{P} W.~Puszkarz, {Higher order modifications of the 
Schr\"odinger equation}, quant-ph/9710007 2 Oct 1997

\end{thebibliography}
\end{document}